\newcommand{\rar}{\rightarrow}
\newcommand{\barr}{\begin{array}}
\newcommand{\earr}{\end{array}}
\newcommand{\bea}[1]{\begin{eqnarray} \label{(#1)}}
\newcommand{\eea}{\end{eqnarray}}
\newcommand{\beq}[1]{\begin{equation} \label{(#1)}}
\newcommand{\eeq}{\end{equation}}

\newcommand{\rarr}{\longrightarrow}

\documentclass[dvips]{article}
\usepackage{icrctc07}

\title{TeV gamma-rays from photo-disintegration/de-excitation of nuclei in Westerlund 2}
\shorttitle{TeV gamma-rays from photo-disintegration/de-excitation of nuclei in Westerlund 2}

\authors{Luis A. Anchordoqui$^1$, John F. Beacom$^2$, 
Yousaf M. Butt$^3$, Haim Goldberg$^4$,\\ Sergio Palomares-Ruiz$^5$, 
Thomas J. Weiler$^6$, and Justin Wesolowski$^1$}
\shortauthors{Anchordoqui et al.}
\afiliations{$^1$Department of Physics, University of Wisconsin-Milwaukee,
Milwaukee, WI 53201\\
$^2$Departments of Physics and Astronomy,
The Ohio State University,
Columbus, OH 43210\\
$^3$Harvard-Smithsonian Center for Astrophysics, 60 Garden St., Cambridge, MA02138\\
$^4$Department of Physics,
Northeastern University, Boston, MA 02115\\
$^5$IPPP, Department of Physics, 
University of Durham, DH1 3LE UK\\
$^6$Department of Physics and Astronomy,
Vanderbilt University, Nashville, TN 37235}
\email{t.weiler@vanderbilt.edu}

\abstract{TeV gamma-rays can result from the photo-de-excitation of
  PeV cosmic ray nuclei after their parents have undergone
  photo-disintegration in an environment of ultraviolet photons. This
  process is proposed as a candidate explanation of the recently
  discovered HESS source at the edge of Westerlund 2. The UV
  background is provided by Lyman-alpha emission within the rich O and
  B stellar environment.  The HESS flux results if there is efficient
  acceleration at the source of lower energy nuclei. The requirement
  that the Lorentz-boosted ultraviolet photons reach the Giant Dipole
  resonant energy ($\sim 20$~MeV) implies a strong suppression of the
  gamma-ray spectrum compared to an $E_\gamma^{-2}$ behavior at energies
  below about $1$~TeV. This suppression is not apparent in the lowest-energy Westerlund 2 datum, 
  but will be probed by the upcoming GLAST
  mission.}

\begin{document}
\maketitle
%Begin the section.

Two well-known mechanisms for generating TeV $\gamma$-rays in
astrophysical sources are the purely electromagnetic (EM) synchrotron
emission and inverse Compton scattering, and the hadronic (PION) one
in which $\gamma$-rays originate from $\pi^0$ production and decay.
Very recently, we highlighted a third dynamic which leads to TeV
$\gamma$-rays: photo-disintegration of high-energy nuclei, followed by
immediate photo-emission from the excited daughter
nuclei~\cite{Anchordoqui:2006pd}.  For brevity, we label the
photo-nuclear process $A+\gamma\rar A'^*+X$, followed by $A'^*\rarr A'
+\gamma$-ray as ``$A^*$''. Such a process may be operative in
massive star formation regions with hot starlight. In this work we
examine whether the $A^*$-process could be the origin of the very
energetic $\gamma$-rays recently observed, with the High Energy
Stereoscopic System (HESS) of Atmospheric Cherenkov Telescopes,
from the young stellar cluster
Westerlund~2~\cite{Aharonian:2007qf}.

RCW 49 is a luminous cloud of ionized hydrogen located towards the
outer edge of the Carina arm, at a distance $d \approx
8$~kpc~\cite{Rauw:2006dr}.  Embedded in RCW 49 is the massive star
formation region Westerlund 2, hosting an extraordinary ensemble of
hot OB stars; presumably at least a dozen early-type O stars, 100~B
stars, and the remarkable Wolf-Rayet binary WR 20a~\cite{Bonanos:2004cb}.  
For such a
distance, the cluster core $\sim 5'$ results in a physical extent $R
\sim 6$~pc. The total mass of the stars within this region is found to
be $\approx 4500~M_\odot .$ The total wind luminosity of all these O
type stars has been estimated as $\sim 5 \times 10^{37}$~erg s$^{-1}$,
and the stellar luminosity of known massive stars is $L = 2.15 \times
10^{40}$~erg s$^{-1}~$\cite{Bednarek}. However, radio emission
from the prominent giant HII region RCW 49 requires a larger
luminosity of ionizing UV photons~\cite{Whiteoak}. Indeed observations
using the Infrared Array Camera (IRAC) on board the Spitzer Space
Telescope indicate that the total number of young stellar objects in
this region is about 7000~\cite{Whitney:2004gq}. Therefore, the above
estimate of the total stellar luminosity should be taken as a lower
bound.

In the vicinity of WR20a, a clear excess of very high energy
$\gamma$-rays was recently reported by the H.E.S.S.
Collaboration~\cite{Aharonian:2007qf}.  The significance of the excess
is about $9\sigma$. Compared to the point spread function of the
instrument, the source (termed HESS J1023--2013575) appears slightly
extended, corresponding to an intrinsic size of the $\gamma$-ray
source of about $0.2^\circ$; its center is slightly shifted compared
to WR 20a. As expected for an extended source, the $\gamma$-ray flux
is steady over time.

By repeating the discussion of Cygnus OB2~\cite{Anchordoqui:2006pe},
in what follows we obtain the expected $\gamma$-ray production through
the $A^*$-process in Westerlund 2.  To compute the
photo-disintegration rate of a highly relativistic nucleus (with
energy $E=A\,E_N=\gamma A m_N,$ where $\gamma$ is the Lorentz factor) on
starlight per nucleon~\cite{Stecker:1969fw},
\begin{equation}
R_A(E_N)=\frac{c}{\lambda_A}  \approx  \frac{\pi \, \sigma_0 \,\epsilon'_0\, \Gamma}{4 \gamma^2}
\int_{\epsilon'_0/2 \gamma}^\infty \frac{d\epsilon}{\epsilon^2}\,\,
  n(\epsilon) \,,
\label{kk}
\end{equation}
we must estimate the ambient photon distribution with energy 
spectrum $n(\epsilon)$. In Eq.~(\ref{kk}) we have approximated 
the Giant Dipole resonant cross section by the single pole of the
Narrow-Width Approximation, 
\begin{equation}
\sigma_A(\epsilon') = \pi\,\,\sigma_0\,\,  \frac{\Gamma}{2} \,\,
\delta(\epsilon' - \epsilon'_0)\, .
\label{sigma}
\end{equation}
Here, $\epsilon'$ is the photon energy in the rest frame of the
nucleus, $\sigma_0/A = 1.45\times 10^{-27} {\rm cm}^2$, $\Gamma =
8~{\rm MeV}$, and $\epsilon'_0 = 42.65 A^{-0.21} \, (0.925
A^{2.433})~{\rm MeV},$ for $A > 4$ ($A\leq 4$). The ambient photon
distribution originates in the thermal emission of the stars in the
core region of radius $R$.  The average density in the region $R$ will
reflect both the temperatures $T_{\rm O}$ and $T_{\rm B}$ due to
emission from O and B stars, respectively, and the dilution resulting
from inverse square law considerations. Specifically, the photon
density is
\begin{equation}
n(\epsilon) = \frac{9}{4} 
\left[\frac{n_{\rm O} (\epsilon) 
N_{\rm O} R_{\rm O}^2 + n_{\rm B}(\epsilon) 
N_{\rm B} R_{\rm B}^2}{R^2} \right],
\label{Jack}
\end{equation}
where $N_{\rm O\,(B)}$ is the number of O~(B) stars , $R_{\rm O(B)}$ is the O(B) star average radius, 
the factor 9/4 emerges when averaging the inverse square
distance of an observer from uniformly distributed sources in a region
$R$, and
\begin{equation}
n_{\rm O(B)} (\epsilon) = (\epsilon/\pi)^2\
\left[e^{\epsilon/T_{\rm O(B)}}-1 \right]^{-1} \,,
\label{nBE}
\end{equation}
corresponding to a Bose-Einstein distribution with temperature $T_{\rm
  O(B)}$.

\begin{figure}
\begin{center}
%\noindent \fbox{\hbox{\vbox{\hsize=130mm \hfill \vspace{50mm}}}}
\includegraphics [width=0.5\textwidth]{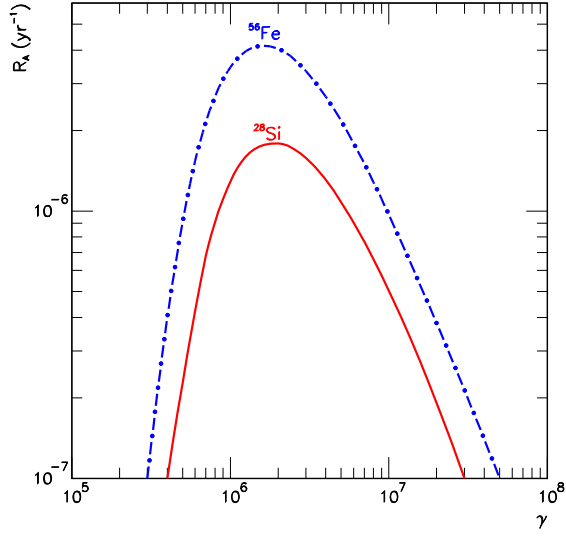}
\end{center}
\caption{Photo-disintegration rates of $^{56}$Fe and $^{28}$Si,
  on the Westerlund 2 starlight.}
\label{Rstar}
\end{figure}

In Fig.~\ref{Rstar} we show the dependence on the Lorentz factor of
$R_{56}$ and $R_{28}$ for the stellar ambiance described above.  We
have taken for the O stars, $N_{\rm O}=12$, a surface temperature $T_O$ = 40000~K, and
radius $R_O = 19\ R_\odot$; for the cooler B stars we assign $T_B$ =
18000~K, $N_{\rm B}=100$, and radius $R_B = 8\ R_\odot$.

The low energy cutoff on $R_A$ seen in Fig.~\ref{Rstar} will
be mirrored in the resulting photon distribution. The $\sim E^{-2}$
energy behavior of the various nuclear fluxes will not
substantially affect this low energy feature.  
The energy behavior for photons in the
$0.5-10$~TeV region of the HESS data is a complex convolution of
the energy distributions of the various nuclei participating in
the photo-disintegration, with the rate factors appropriate to the
eV photon density for the various stellar populations.

Let us define the differential rate $dR_A/dE'_\gamma$ as
\begin{eqnarray}
\frac{dR_A}{dE'_\gamma} & = & \frac{1}{2} \, \int_0^{\infty}
\frac{n(\epsilon)}{\gamma^2 \epsilon^2}  \, d\epsilon \nonumber \\
 & \times &
\int_0^{2\gamma \epsilon} \epsilon' \,
\frac{d\sigma_{\gamma A}}{dE'_\gamma}(\epsilon',E'_\gamma) \,
d\epsilon' \,\,,
\end{eqnarray}
where $d\sigma_{\gamma A} (\epsilon',E'_\gamma)/dE'_\gamma$ is the
inclusive differential cross section for production of $\gamma$-rays
from disintegration and $E'_\gamma$ is the energy of the emitted
photon(s) in the rest frame of the nucleus. Assuming the same cosmic ray
spectrum as above, the emissivity (number/volume/steradian)
of $\gamma$-rays coming from nuclei photo-emission is 
\begin{eqnarray}
Q_\gamma^{\rm dis} (E_\gamma) &  = & \sum_A \int \frac{dn_A}{dE_N}(E_N) \,
dE_N \nonumber \\
& \times & \int  \frac{dR_A}{dE'_\gamma} 
\, dE'_\gamma \,
\frac{d\cos{\theta_\gamma}}{2} \nonumber \\
 & \times & \delta[E_\gamma - \gamma
  E'_\gamma (1+\cos{\theta_\gamma})] \,\,,
\end{eqnarray}
where $E_\gamma$ is the energy of the emitted $\gamma$-ray in the lab and
$\theta_\gamma$ is the $\gamma$-ray angle with respect to the
 direction of the excited nucleus.
Assuming a power law with spectral index $\alpha$ for the nuclear flux,
and approximating the $\gamma$-ray spectrum as being monochromatic with
energy  $\overline{E'_{\gamma A}}$,
the emissivity becomes~\cite{Karakula:1994nv}
\begin{eqnarray}
\label{qgdis}
Q_\gamma^{\rm dis}(E_\gamma) & = & \sum_A \frac{\overline{n_A} m_N}{2
  \overline{E'_{\gamma A}}} \int_{\frac{m_N E_{\gamma
  }}{2\overline{E'_{\gamma A}}}}   \frac{dE_N}{E_N} 
 \nonumber \\
 & \times & R_A(E_N) \, \frac{dn_A}{dE_N}(E_N)\, ,
\end{eqnarray}
where $\overline{n_A}$ is the mean
$\gamma$-ray multiplicity for a nucleus with atomic number $A$. 
(Hereafter we take $\overline{n_A}=2$).
The differential flux at the
observer's site (assuming there is no absorption)  is related to to
the $\gamma$-ray emissivity as
\begin{equation}
\label{fluxg}
\frac{dF_\gamma}{dE_\gamma} (E_\gamma) =
\frac{V_{\rm dis}}{4 \pi d^2} \, Q_\gamma^{\rm dis} (E_\gamma) \,,
\end{equation}
where $V_{\rm dis}$ is the volume of the source (disintegration) region
and $d$ is the distance to the observer.

\begin{figure}
\begin{center}
\includegraphics [width=0.5\textwidth]{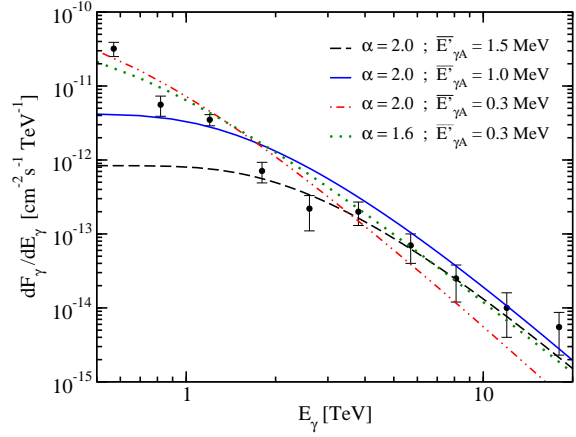}
\end{center}
\caption{Sample eyeball fits to HESS Westerlund 2 $\gamma$-ray
  spectrum for the $A^*$~model, with various values of the injection
  $^{56}$Fe nuclei power index $\alpha$ and the average energy
  $\overline{E'_{\gamma A}}$ of emitted de-excitation photons.}
\label{specs}
\end{figure}

In Fig.~\ref{specs} we provide a sample of eyeball fits to the gamma
ray spectrum, as obtained following a direct integration of
Eq.~(\ref{qgdis}).  The fits are for interesting choices of
the spectral index ($\alpha$) of the nuclei population and for the average
energy of the photon (in the nuclear rest frame) emitted during
photo-emission.  It is apparent that the A* mechanism can provide
reasonable agreement with the data, except possibly for the
lowest-energy datum.  This datum may require an alternate mechanism,
such as electron acceleration or a PION
contribution~\cite{Butt:2007rr}.

The EM and PION~$pp$ processes contrast with the $A^*$-process in that
for them there is either no energy threshold (EM) or very small
threshold ${\cal O} (2m_\pi)$ (PION~$pp$) in the lab, and so their
resulting $\gamma$-ray spectra rise monotonically with decreasing
$E_\gamma$.  In contrast, the $A^*$ spectrum is flat with decreasing
$E_\gamma$ below about a TeV, as is evident in Fig.~\ref{specs}.

The conversion efficiency from cosmic-rays to TeV~$\gamma$-rays must
be quite high in the $A^*$ model, as with other models.  We find that
the ratio of power in the nuclei flux to that in the O-star winds is
\begin{eqnarray}
\label{efficiency}
\frac{P_A}{P_{\rm O\;wind}} & = & {\rm eff}\times \left(\frac{R}{6\,{\rm pc}}\right)^2\,
\left(\frac{d}{8\,{\rm kpc}}\right)^2  \nonumber \\
&  \times & \left(\frac{40}{N_{\rm O}}\right)^2\,
\left(\frac{\tau_{\rm age}}{\tau_{\rm cont}}\right)\,
\end{eqnarray}
where ${\rm eff}$ is an efficiency which depends on model parameters,
$\tau_{\rm age}\sim 2\,{\rm Myr}$ is the age of the star-forming
region~\cite{Bednarek}, and $\tau_{\rm cont}$ is
the containment time of the nuclei in the ambient magnetic field.  For
the four parameter sets $(\alpha,\overline{E'_{\gamma A}}/{\rm MeV})$
listed in Fig.~\ref{specs}, we find for the (2.0, 1.5), (2.0, 1.0),
(1.6, 0.3) and (2.0, 0.3) models, the respective efficiency values of
${\rm eff}=4\%$, 8\%, 20\%, and 80\%.\\

In summary, we have shown that the observed TeV $\gamma$-rays from
Westerlund 2 can be explained by the $A^*$~model, wherein the TeV
$\gamma$-rays are the Lorentz-boosted MeV $\gamma$-rays emitted on the
de-excitation of daughter nuclei, themselves produced in collisions of
PeV nuclei with a hot ultraviolet photon background.  There is a
specific prediction of a suppression of the $\gamma$-ray spectrum in
the region below $1~{\rm TeV}$ -- this because of the need to achieve
Giant Dipole Resonance excitation through collision with $\sim$ few eV
photons.  This suppression is not apparent in the lowest-energy
Westerlund 2 datum, but will be probed by the upcoming GLAST
mission~\cite{Gehrels:1999ri}: The flux predicted at $\sim$ 100 GeV
from an $E_\gamma^{-2.53}$ extrapolation of the HESS
data~\cite{Aharonian:2007qf} would render the source spectacularly
visible in the GLAST observation, whereas the $A^*$-model predicts a
suppression by a factor of more than 2 orders of magnitude relative to
this extrapolated flux.

We have also calculated the energy requirements for the proposed mechanism
by integrating over the nuclei energy density. We have found that, in
order to fall within the estimated kinetic energy budget of the O~stars, the
containment time needs to be large (of ${\cal O}(10^5 \rm{yr})$), so 
that the wind power integrates in time to a sufficiently large
energy.  
It should be noted that WR~20a may by itself contribute several times the 
wind energy of the O~stars, thereby lowering our effeciency factor by a comparable amount.\\

The research of JFB is supported by The Ohio State University and the
NSF CAREER Grant No.  PHY-0547102.  YMB is supported by NASA/Chandra
and NASA/INTEGRAL GO Grants and a NASA LTSA Grant.  HG is supported by
the U.S. NSF Grant No PHY-0244507.  SPR is partially supported by the
Spanish Grant FPA2005-01678 of the MCT.  TJW is supported by the U.S.
DOE Grant No. DE-FG05-85ER40226.  HG, SPR, and TJW thank the Aspen
Center for Physics for a productive environment during this work.

\end{document}